# *In-materia* speech recognition

Mohamadreza Zolfagharinejad[1], Julian Büchel[2], Lorenzo Cassola[1], Sachin Kinge[3],

Ghazi Sarwat Syed[2], Abu Sebastian[2], Wilfred G. van der Wiel[1,4†]

[1]NanoElectronics Group, MESA+ Institute for Nanotechnology and BRAINS Center for

Brain-Inspired Nano Systems, University of Twente,

PO Box 217, Enschede 7500 AE, The Netherlands.

[2]IBM Research - Europe, Säumerstrasse 4, 8803 Rüschlikon, Switzerland

[3]Materials Research and Development, Toyota Motor Europe, B-1930 Zaventem, Belgium

[4]Institure of Physics, University of Münster, 48149 Münster, Germany

[†]Correspondence to: W.G.vanderWiel@utwente.nl

**With the rise of decentralized computing, as in the Internet of Things, autonomous driving, and personalized healthcare, it is increasingly important to process time-dependent signals 'at the edge' efficiently: right at the place where the temporal data are collected, avoiding time-consuming, insecure, and costly communication with a centralized computing facility (or 'cloud'). However, modern-day processors often cannot meet the restrained power and time budgets of edge systems because of intrinsic limitations imposed by their architecture (von Neumann bottleneck) or domain conversions (analogue-to-digital and time-to-frequency). Here, we propose an edge temporal-signal processor based on two *in-materia* computing systems for both feature extraction and classification, reaching near-software accuracy for the TI-46-Word[1] and Google Speech Commands[2] datasets. First, a nonlinear, room-temperature dopant-network-processing-unit (DNPU) [3,4] layer realizes analogue, time-domain feature extraction from the raw audio signals, similar to the human cochlea. Second, an analogue in-memory computing (AIMC) chip[5], consisting of memristive crossbar arrays, implements a compact neural network trained on the extracted features for classification. With the DNPU feature extraction consuming ~300 nJ and AIMC-based classifier ~78 μJ with sub-millisecond latency and the potential for less than 10 fJ per multiply-accumulate operation[6], our findings offer a promising avenue for advancing the compactness, efficiency, and performance of heterogeneous smart edge processors through *in-materia* computing hardware.**

Evolution has endowed the human ear with an ingenious sound preprocessing system that performs two domain conversions. Hair cells on the basilar membrane in the cochlea decompose sound waves into their frequency components (time-to-frequency conversion) and convert their mechanical vibrations into electrical signals (mechanical-to-electrical conversion) [7], which are transmitted to the brain by the auditory nerve for cognitive recognition and interpretation (Fig. 1a) [8]. The auditory system, with its active subsystems, not only converts acoustic waveforms into neural spikes but also *generates* additional tones[9]. As the hair cell vibrations do not increase linearly with the sound intensity and the cochlea exhibits an active feedback mechanism, non-harmonic frequencies (distortions) are generated (Fig. 1a). This *compressive* nonlinearity is crucial for the sensitivity, frequency selectivity, and dynamic range in hearing[9]. Thus, the in-ear preprocessing thus provides real-time, low-power encoding, significantly reducing the data needed to represent raw auditory information.

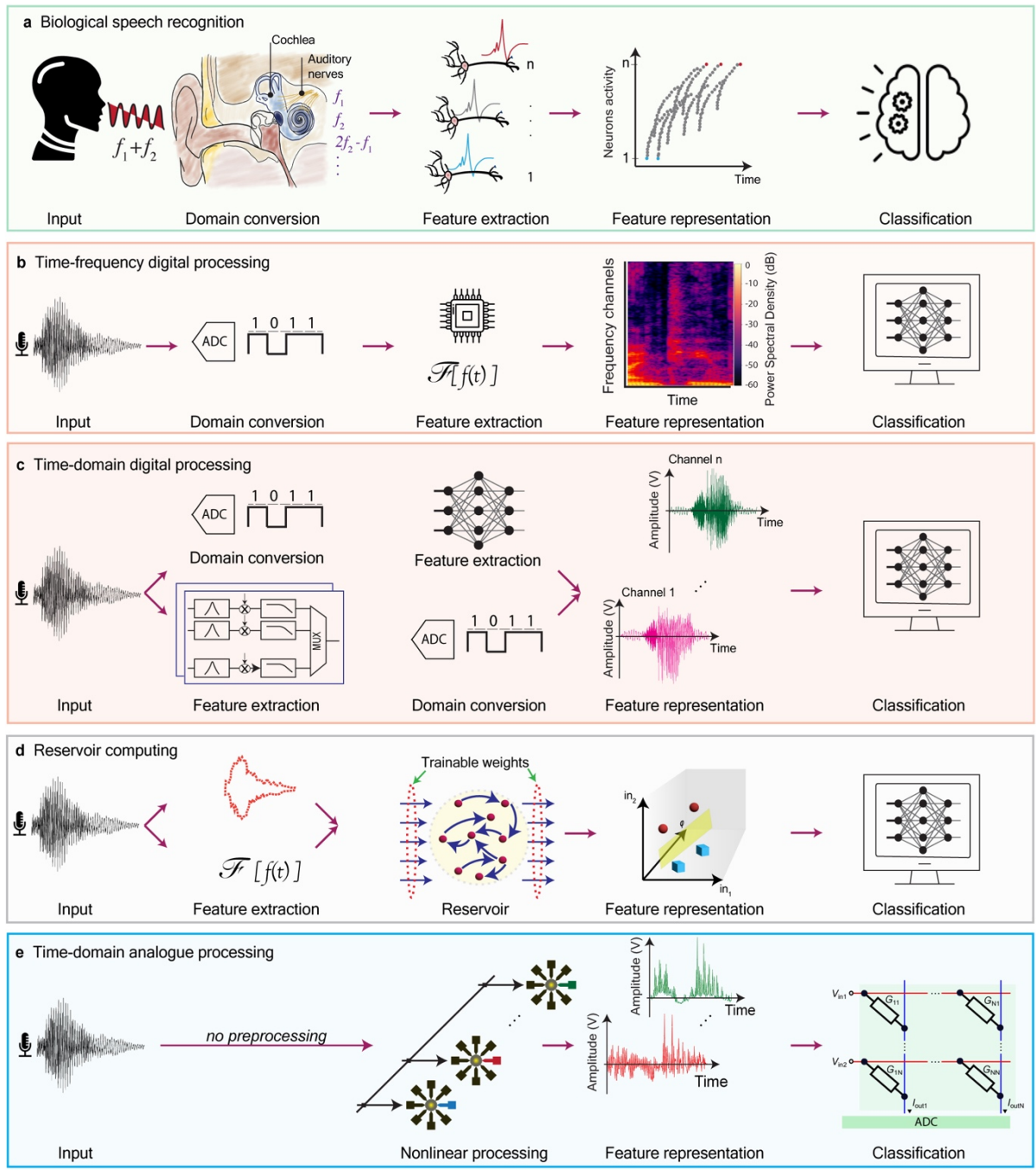

**Figure 1. Overview of speech recognition approaches. (a)** Biological speech recognition. As an example, when a sound with two frequencies $f_1$ and $f_2$ ($f_2 > f_1$), enters the ear, a family of distortion products, such as $2f_1 - f_2$, is generated due to the nonlinear active feedback system in the cochlea[9,10]. The hair cells on the basilar membrane connected to the auditory nerve endings (1 to *n*) incorporate two domain conversions to convert the incoming time-dependent acoustic signal into electrical spike-encoded frequency-dependent information (features) to be further

processed (classified) by the brain. **(b)** Time-frequency digital processing method. After analogue-to-digital conversion (ADC), frequency decomposition by a feature-extracting model $F\,[f(t)]$, such as Lyon's artificial-cochlea model[11], is required prior to the classification. **(c)** Time-domain digital processing method. Upper part: in addition to classification, a neural network performs feature extraction by learning (bandpass) filters in the time domain. Lower part: an analogue filter bank extracts frequency features directly from the time-dependent, analogue signal prior to the classification. **(d)** Reservoir computing method. After feature extraction in the time (upper part[12]) or frequency domain (lower part), the preprocessed data are projected into a high-dimensional feature space, making the subsequent linear classification simpler. **(e)** Time-domain analogue processing approach (this work). The nonlinear transform provided by dopant network processing units (DNPUs) extracts temporal features and reduces the classification complexity without the need for any additional preprocessing. An analogue in-memory computing (AIMC) chip consisting of a crossbar array of memristive devices is used to efficiently incorporate the classification.

Computerized audio signal processing, particularly automatic speech recognition (ASR), commonly incorporates three domain conversions (Fig. 1b). First, a microphone converts the acoustic waveform into an analogue electrical signal; second, this signal is digitized using an analogue-to-digital converter (ADC); and third, a short-term Fourier transform (STFT) is applied. The time-to-frequency domain conversion is an important part of the *feature extraction* stage, which – potentially in combination with additional nonlinear data transformations – significantly reduces the complexity of the subsequent classification stage[13]. The latter can be accomplished by an artificial neural network (ANN), such as a transformer[14], a convolutional neural network (CNN)[15], a recurrent neural network (RNN)[16], or a combination of those[17,18].

ASR models are typically run on specialized hardware, such as application-specific-integrated-circuit accelerators, like a neural engine[19,20], graphics processing units[21], or tensor processing units[22]. The availability of fast and accurate digital hardware has broadened ASR applications, ranging from virtual assistants and customer service systems to transcription services and language-learning platforms. There is particularly an increasing demand for ASR 'at the edge' in applications such as always-on e-health medical systems[23], offline voice assistants[20], and fault-detection systems[24,25], where latency, privacy concerns, or limited connectivity play a role of importance. These requirements demand real-time (or low-latency) and ultra-low-power computing.

However, digital temporal signal processing at the edge is challenging due to expensive domain conversions, complex feature-extraction algorithms[26] and the von Neumann bottleneck[27]. Although combined feature extraction and classification in the time domain (Fig. 1c, upper part) omits the time-to-frequency domain conversion by porting the feature extraction to a neural network[15,28,29], it generally requires larger neural networks to extract features from raw data[30,31]. These difficulties have led to a growing interest in alternative, more efficient approaches[32-34].

Apart from the conventional (*i.e.*, fully digital) approaches of Figs. 1b and 1c (upper part), emerging ASR can roughly be divided into three categories: artificial-cochlea models in combination with spiking neural networks (SNNs), closest to human hearing (Fig. 1a) [16]; analogue filter banks[35,36] (Fig. 1c, lower part); and reservoir computing (RC, Fig. 1d) [12,37]. Artificial-cochlea implementations perform feature extraction by converting analogue signals into 'spike grams'[38]. Despite the similarity to biology, a spiking hardware classifier realized with, *e.g.*, a digital SNN is required. Lower classification accuracies with spiking features, compared to time-, or frequency-domain features, suggest that the classification step is more difficult[39]. Although analogue filter banks (Fig. 1c, lower part) do not require digitization for feature extraction, recent demonstrations still require digital classifiers[35]. In RC[40-42], the reservoir first projects the temporal input into a higher-dimensional feature space, enabled by its recurrent (multi-timescale) time dynamics, followed by a linear classification layer. However, high classification accuracy requires a complex classification step when the reservoir is fed with time-domain input[12], *e.g.,* a deep CNN[31]. Alternatively, expensive preprocessing, such as using the mel-frequency cepstral coefficients, is needed when feeding the reservoir with frequency-domain features[13,26,37,43-46]. Despite promising classification results obtained on the TI-46-Word spoken digit task[47], edge conditions demand hardware that is not only accurate but also compact and efficient. This requires an approach that simultaneously addresses domain conversions, efficient feature extraction and the von Neumann bottleneck.

In this Article, we present an ASR architecture based on two emerging *in-materia* computing paradigms. First, analogue, time-domain feature extraction is achieved through a circuit that incorporates one or more dopant network processing units (DNPUs) [3] working at room temperature. Serving as the core of a low-power analogue circuit, the DNPU uniquely establishes biologically plausible feature extraction through a recurrent nonlinear transformation with adjustable frequency selectivity and tuneable characteristic time scales in the millisecond range. Second, for the classification based on the extracted features, we implement a CNN on an analogue in-memory computing (AIMC) [5] chip. Crossbar arrays of synaptic phase-change-memory (PCM) unit cells perform the matrix-vector multiplication operations for the CNN, mitigating the von Neumann bottleneck. We assess the performance of these combined *in-materia* approaches using two commonly used benchmarks, namely, the TI-46-Word spoken digits and Google Speech Commands datasets, achieving near-software-equivalent accuracies of 96.2 ± 0.8% and 89.3%, respectively.

**Time-domain processing with DNPUs**

DNPUs possess the three characteristics crucial for biologically plausible time-domain speech recognition: frequency selectivity, tuneable nonlinearity, and sensory memory within the desired millisecond range in audio recognition. Figure 2(a) shows a false-colour atomic-force-microscopy (AFM) image of a DNPU and the measurement circuitry. The active region of the DNPU consists of a disordered network of hopping sites in silicon, surrounded by eight electrodes (one input, one output, six controls). This network features a complex energy landscape and highly nonlinear charge transport tuneable via the control voltages to achieve specific functionalities, such as Boolean logic, nonlinear classification, and feature extraction[3,4,48] (see Methods for fabrication details).

In our earlier work on static (time-independent) tasks[3,4,48], we operated the DNPUs at 77 K with voltage input and current output, using a transimpedance amplifier (Extended Data Fig. 1), effectively shorting the stray capacitance of the readout circuit, and minimizing the response time. Here, for the first time, we operate at room temperature using a modified fabrication process avoiding HF etching, likely enhancing the role of $P_b$ centres. The devices now show ~0.4 eV activation energy and nonlinear behaviour possibly due to a combination of trap-assisted and space-charge-limited-current transport mechanisms (see Methods and Suppl. Info. Note 10). Furthermore, we measure the output as a *voltage*. Specifically, we use a buffer circuit with a large input impedance (> 1 GΩ) at the DNPU output, so that the external capacitance $C_{ext}$ (~10-100 pF) can no longer be neglected (Fig. 2a). Together with the intrinsic nonlinear DNPU resistance and capacitance, these features now give rise to a highly nontrivial transformation with a tuneable characteristic timescale in the ~ms timescale. In the inset of Fig. 2b, we show the distribution of these timescales, corresponding to 500 sets of random control voltages.

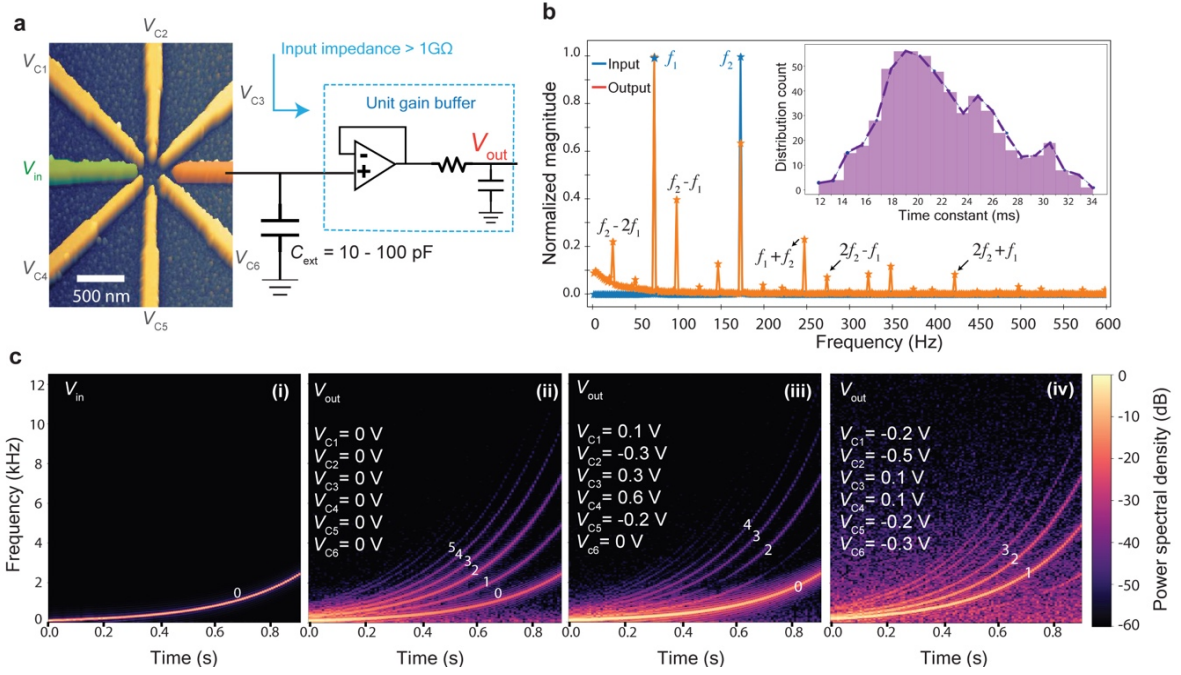

**Figure 2. Tuneable nonlinearity and short-term memory characterization in a DNPU at room temperature. a,** Atomic-force-microscopy image of an 8-electrode DNPU and schematic measurement circuit with external capacitance $C_{ext}$ (~10-100 pF) and a >1 GΩ input impedance buffer. **b,** Fourier transform of a two-tone input signal (blue, $f_1$ and $f_2$) and the DNPU response (orange), showing (sub-)harmonics and distortions. Input and output signals are sampled with a fixed sampling rate of 25 kS/s. Similar to the frequency response of a human cochlea, the output signal contains a family of distortion products ($|(n+1)f_1 - nf_2|$) of progressively higher frequencies and lower magnitudes[9]. Inset: Histogram of time constants ($\tau$ is the time for the output to reach 63% of $V_{max}$) of $V_{out}$ as a response to a 1 V step input with a rising time <20 μs for 500 randomly configured control-voltage sets. **c(i),** Power spectral density (PSD) of chirp input voltage signal $V_{in}(t) = A \sin\left[2\pi f_0 \left(\frac{k^t - 1}{\ln(k)}\right) + \varphi_0\right]$, where $A$ is the amplitude, $f_0$ the starting frequency, $k$ the rate of exponential change in frequency, $\varphi_0$ the initial phase at $t = 0$, $f_1$ the final frequency, and $T$ the duration of the signal. As shown in Fig. 2c(i), these values are set to $A = 0.75$ V, $\varphi_0 = -\frac{\pi}{2}$, $k = (\frac{f_1}{f_0})^{\frac{1}{T}}$, $T = 1$ s, $f_1 = 2$ kHz, and $f_0 = 100$ Hz. **c(ii),** PSD of output voltage $V_{out}(t)$ in response to the input signal of c(i) with zero volts on all controls, showing up to 5 harmonics. **c(iii),** Idem for a set of random control voltages. **c(iv),** Idem, for a different set of random control voltages.

Figure 2b shows the response of the DNPU circuit to a two-tone input, where a signal with two frequencies ($f_1 = 74$ Hz, $f_2 = 174$ Hz) is fed to the DNPU input (blue). The Fourier transform of the output signal is shown in orange, where, by changing the control voltages, the magnitude of these distortion products can be adjusted. Figures 2c(ii)-(iv) show the output spectrograms for three different sets of control voltages in response to an exponential chirp input signal (Fig. 2c(i)). In Fig. 2c(ii), all control electrodes are set to 0 V with respect to the ground. The output voltage contains up to the 5$^{th}$ harmonic of the input signal, indicating the presence of strong nonlinearities. Importantly, these harmonics can be tuned by the control voltages, as shown in Figs. 2c(iii) and 2c(iv). For example, harmonics can be selectively created or removed, including the fundamental tone.

The DNPU output voltage is determined by the complex potential landscape of the active region, which in turn depends on the potential at each electrode. Similar to a state-dependent system, when the charge stored at the output capacitor $C_{ext}$ changes, the potential at the output electrode correspondingly alters, modifying the circuit characteristics (*i.e.,* $\tau$), effectively forming an active-feedback system, which introduces recurrency and short-term memory (Fig. 3b). In concrete, the charge on $C_{ext}$ at time $t = t_0$ influences the DNPU behaviour at $t = t_0 + \Delta t$, given $\Delta t \lesssim R_{DNPU}(t_0) C_{ext}$, where $R_{DNPU}(t_0)$ is the DNPU resistance measured at $t = t_0$ between the input and output electrodes with applying the control voltages $V_{C1}$, $V_{C2}$, …$V_{C6}$. Supplementary Info. Note 1 provides details on the recurrent form of fading memory based on the variable impedance of the DNPU.

Frequency selectivity and short-term (fading) memory are key characteristics of neural network models for temporal data processing[12,15]. For example, CNN models use large kernel sizes for the first convolution layer, typically a ~10 ms receptive field, to capture temporal features from raw audio[29,30]. The first layer extracts low-level features, while the deeper layers with smaller kernel sizes construct higher-level features and perform classification[29]. Interestingly, the overall accuracy of an end-to-end CNN classifier is only slightly affected by training the initial layer and remains acceptable even with random initializations[15]. Leaving the first layer untrained, *i.e.,* feature extraction with random software filter banks[49] can reduce the training costs[50]. In the inference phase, however, the first layer is the most computationally expensive part of the network as the large kernel size demands many multiply-accumulate (MAC) operations. For example, in a 5-layer CNN for raw audio classification, for inference of a single input sample, the first layer accounts for ~75% of all MAC operations[30] (Suppl. Info. Note 2).

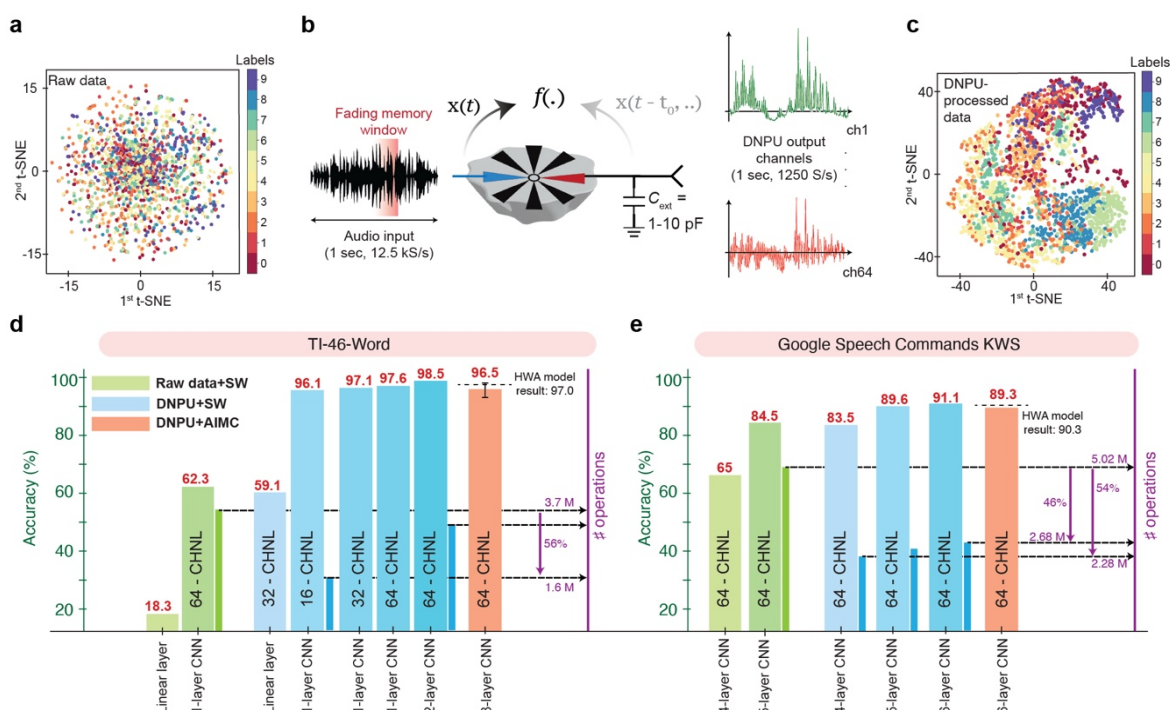

**Figure 3. DNPU analogue feature extraction for speech recognition. a,** t-distributed stochastic neighbour encoding (t-SNE) visualization for the female subset of the TI-46-Word spoken digit dataset before DNPU preprocessing. **b,** Schematic of DNPU (nonlinear function noted as $f(.)$) fed with analogue time-dependent input $x(t)$ (blue electrode), voltage measured at the orange electrode and constant control voltages (black electrodes). Every set of control voltages results in a unique transformed signal (green and red curves shown as examples) and forms an output channel with 10× lower sample rate compared to the raw input signal (see Methods). **c,** t-SNE visualization after DNPU preprocessing (one configuration out of 32 sets of randomly chosen control voltages). The output data show that the DNPU preprocessing helps clustering utterances of the same digit, simplifying later classification. **d,** Comparison of the TI-46-Word classification accuracy for linear and CNN classifier models without (green, software model) and with (blue, software model and red, hardware-aware (HWA) trained model) DNPU preprocessing for 16, 32, and 64 DNPU channels. The all-hardware (DNPU with the AIMC classifier) result (orange) is presented as the mean ± one standard deviation over ten inference measurements. **e,** Comparison of the Google Speech Commands (GSC) keyword spotting (KWS) classification accuracy with/without DNPU preprocessing with 4, 5, and 6-layer CNNs. DNPU preprocessing allows for achieving higher accuracies with classifiers requiring over twice fewer multiply-accumulate operations.

**Speech recognition and keyword spotting with DNPU feature extraction and software classifiers**

Motivated by insights from biology and conventional speech-recognition systems, we propose a DNPU circuit with its *in-situ* adjustable nonlinearity (Figs. 2b and 2c) and tuneable time dynamics (Fig. 2b, inset) to implement an *in-materia* time-domain feature extractor and combine it first with a software classifier model (we discuss an *in-materia* classifier below). To benchmark the performance of this hybrid system, we use the TI-46-Word spoken digits and a subset of the Google Speech Commands (GSC) datasets (see Methods). These datasets are processed 64 times by a single DNPU circuit, each process step referred to as a *channel* (Fig. 3b) and with a different set of random control voltages (in a device-dependent voltage range, see Methods). Alternatively, one could obtain 64 channels by using 64 different, randomly initialized DNPUs in parallel. Also, instead of randomly initialized DNPUs, one could use DNPUs with control voltages *trained* in an end-to-end fashion together with the software classifier, using backpropagation (see Suppl. Info. Note 3).

To visualize the impact of DNPU preprocessing, we use t-distributed stochastic neighbour encoding (t-SNE)[13], which is a statistical method for visualizing high-dimensional data[51] (we also used universal manifold approximation and projection (UMAP)[52], see Suppl. Info. Note 4). Figure 3a shows the t-SNE visualization of the original (raw) TI-46-Word audio dataset. Every individual sample is fed to the circuit shown in Fig. 3b and the output is digitized and recorded. As one of the output channels is visualized in Fig. 3c, compared to the original dataset, well-defined clusters are observed for some digits, such as 'six', 'eight', and 'nine', whereas, for other digits, the distribution is more scattered. The results of the DNPU preprocessing (illustrated by the green and red traces in Fig. 3b) contain fingerprints (or low-level features) of the input data. Similar to contrastive learning[53], the DNPU preprocessing puts same-labelled instances of the dataset closer together in *time-domain* representation space while keeping dissimilar instances apart.

To study the impact of DNPU preprocessing on the overall classification performance, we examined several simple classifier models, with and without DNPU preprocessing (see Methods). As shown for TI-46-Word in Fig. 3d, a single-layer linear model performs poorly on the raw data (green, ~18%). An additional single convolutional layer with 32 output channels after, or 32-channel DNPU preprocessing (blue) before the linear model improves the overall classification accuracy to ~60%. It is worth noting that the backpropagation-trained software CNN outperforms preprocessing by *randomly* initialized DNPUs by only 3% point. If we combine 16-channel DNPU preprocessing with a 1-layer CNN, we achieve 95.1% accuracy (blue), with only ~4,000 learnable parameters. For comparison, recent studies on a subset of the TI-46-Word spoken digits dataset with 500 samples, using the same number of learnable parameters (4,096), have reported classification accuracies of 78% without[12] and 97% with[54] frequency-domain feature extraction. We obtain higher accuracies with 32 or 64 channels and using a 2- or 3-layer CNN software model. With 64-channel DNPU-preprocessed data and a 2- (and also 3-) layer CNN, we now obtain 98.5% overall accuracy (blue), which is comparable to the best known all-software solutions, such as long short-term memory (LSTM) models[54].

DNPU preprocessing similarly enhances the classification accuracy for complex keyword spotting (KWS) tasks using the GSC dataset while reducing the required operations by more than a factor of 2 (Fig. 3e). A 5-layer CNN, where the first layer has a convolution layer with a ~10 ms receptive field[28], requires ~5M MAC operations and gives 84.5% classification accuracy for the same subset of the GSC KWS task. In comparison, with 64-channel DNPU preprocessing, a 4-layer CNN that only requires 2.28M MAC operations achieves a comparable 83.5% classification accuracy. The accuracy increases to 91.1%, close to the all-software, state-of-the-art value of 92.4%, with DNPU-preprocessed data and a compact 6-layer CNN. DNPU preprocessing combined with the same CNN architecture using the Swish activation function[55] instead of hyperbolic tangent, achieves the highest classification accuracy of 93.1%.

*Frequency selectivity, compressive nonlinearity*, and a *recurrent* form of feedback from the output are key properties of the DNPU circuit that facilitate temporal data processing. These characteristics are indispensable elements for acoustic feature extraction, which are also fundamental in the biological cochlea. To substantiate their importance, in Suppl. Info. Notes 5-7, we compare different filter banks of low- and bandpass filters when they incorporate one or more of the mentioned characteristics. There, we show that having, for instance, a *hyperbolic tangent* compressive nonlinearity can notably improve the feature extraction performance in filter banks. Furthermore, in Suppl. Info. Note 8, we use 64 different *reservoir* models for the feature extraction step, where we can see that although high-level projections of reservoirs are, to some extent, capable of acoustic feature extraction, they perform poorly compared to other methods. Supplementary Table I summarizes the inference accuracy among different preprocessing methods for the same classifier model, indicating the superior performance of DNPUs.

**Mapping the classifier model on an AIMC chip**

As discussed above, DNPU preprocessing effectively extracts time-domain acoustic features that significantly simplify the classification task. However, the classifier model still consumes typically ~40% of the total power consumption in an ASR system[35]. In traditional DNN implementations, the energy consumption is dominated by memory access and data transfer between the processor and memory units[56]. We leverage analogue in-memory computing or AIMC, another *in-materia* computing paradigm, to address this challenge. Memristive materials provide processing *in* non-volatile memories to store the weights of the neural network[33,57,58]. Below, we discuss the results of a classifier model trained with the features from the DNPU implemented on the IBM HERMES project AIMC chip[5] (Fig. 4a, through a digital interface as the DNPU and AIMC chips are currently physically separated).

First, we implemented the CNN model illustrated in Fig. 4b on the HERMES project AIMC chip for TI-46-Word classification, which is trained with 64 DNPU measurement channels. The training includes a standard and a hardware-aware (HWA) retraining phase to enhance the network model's resilience to analogue noise[59]. We used the open-source IBM analogue hardware acceleration kit[60,61] to add weight and activation noise, perform weight clipping, and incorporate ADC quantization noise during the retraining phase (see Methods for more details). The learned network weights are then transferred to each synaptic PCM unit cell as analogue conductance values (illustrated in Fig. 4a) using gradient-descent-based programming (GDP)[62]. The resource utilization of this model is shown in Fig.

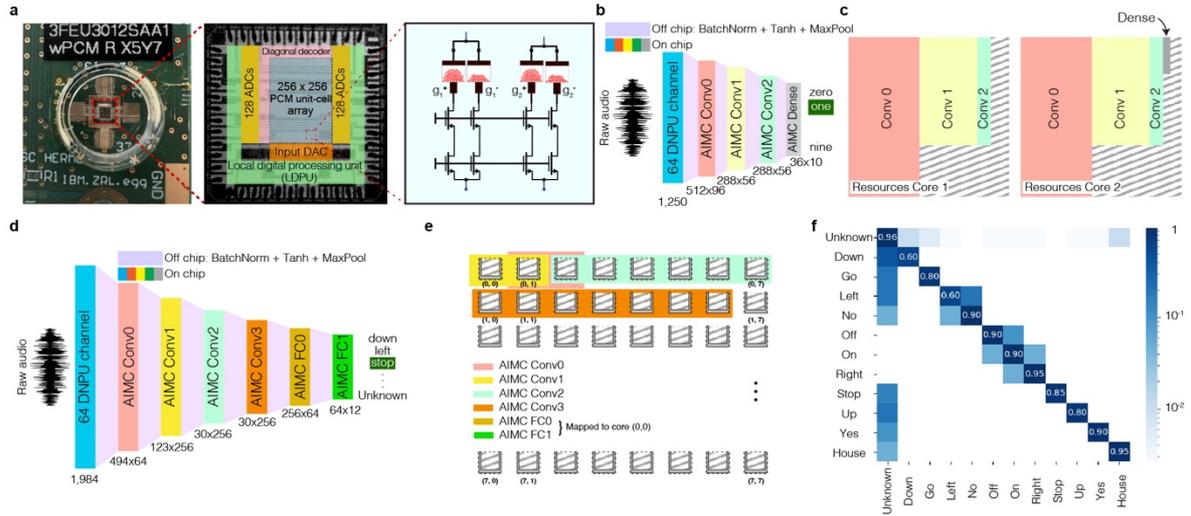

**Figure 4. Schematic of a hybrid convolutional neural network (CNN) architecture for *in-materia* speech recognition. a**, A photograph of the IBM HERMES project chip and its architecture containing 256 × 256 synaptic unit cells, each comprising of 4 phase-change-memory (PCM) devices organized in a differential configuration, ADC/DAC arrays, and on-chip local digital processing units (LDPUs). **b,** Classifier architecture for the TI-46-Word dataset. A 64-channel DNPU preprocessing step converts audio signals into 64-D input to the AIMC with a down-sampling rate of 10 (Details in Extended Data Figure 3). Batch normalization, activation functions, and pooling operations are performed off-chip. **c**, Schematic representation of resource utilization for the 3-layer CNN classifier of panel b for the TI-46-Word implemented on two tiles of the AIMC chip. **d,** CNN architecture for the GSC KWS task. After 64-channel DNPU pre-processing, a 6-layer CNN maps the inputs into 11 classes of known or one class of unknown targets. **e,** The mapping of CNN layers on the AIMC chip. In a fully pipelined implementation, 18 cores (out of 64) of the Hermes chip will be utilized. **f,** The confusion matrix of the GSC KWS task with HWA model training. True-positive rate of the unknown class reaches ~96% accuracy with the overall accuracy being ~91%.

4c. We utilize only a fraction of two cores out of the chip's sixty-four, highlighting the compactness of the model. The two combined *in-materia* systems achieve a near-software inference accuracy (96.2 ± 0.8% averaged over 10 repetitions, compared to 98.5% for the DNPU with a software classifier) while performing approximately 95% of total operations on novel material systems and offloading less than 5% to digital processors.

Similarly, Fig. 4d illustrates a 6-layer CNN for the GSC KWS task implemented on the AIMC chip with its resource mapping shown in Fig. 4e. Although the AIMC chip has been measured through time-multiplexing, a fully pipelined implementation, where every layer is realized by a unique in-memory computing core, requires only 18 cores (out of 64), *i.e.,* less than 29% resource utilization. This small footprint implies that an optimized AIMC chip can be much more compact. The confusion matrix (Fig. 4f) indicates that the "unknown" class is the most challenging, where multiple utterances, such as numbers from zero to nine, have to be linked to one label, *i.e.*, "unknown". The true-positive rate for this class is 96%, while the overall accuracy across all labels reaches 91%.

**Discussion**

To the best of our knowledge, this is the first demonstration of bio-plausible, time-domain speech recognition using two *in-materia* computing hardware systems for both feature extraction and classification. The 96.2% and 89.3% end-to-end, all-hardware accuracies (*i.e.*, DNPU preprocessing + AIMC classification) obtained for the TI-46-Word and GSC KWS tasks in this study are comparable to those of state-of-the-art software models that have ~10× more learnable parameters[54], and only, on average, ~2% point lower than when the classifier is implemented in software (*i.e.*, DNPU + software classification). End-to-end training with DNPUs (Suppl. Note. 3) – rather than leaving them untrained – as well as optimizing the AIMC HWA retraining phase, specifically for temporal data processing, can further improve these results. Furthermore, in contrast to the state of the art, our approach achieves high classification rates without the necessity for costly digital feature extraction[43] or a complex classifier model[31].

We have evaluated the overall system-level efficiency, considering a raw audio waveform as input and classification label as output (Suppl. Note. 9). The analysed system comprises DNPU feature extraction circuits, a digital interface with ADCs[35,63] (or equivalently an analogue interface with sample-and-hold circuits), and the IBM HERMES project AIMC classifier chip. Among feature extraction methods, MFCC features exhibit similar nonlinearity to DNPUs, simplifying the classification step[13]. However, their state-of-the-art digital implementation requires ~10 µW of power and ~0.5 mm² of silicon area with a latency of >10 ms[64]. In contrast, our DNPU feature-extraction circuit consumes ~300 nW (>30× lower), uses ~1 µm² of silicon per DNPU channel (~100 µm² if external capacitance is needed) and operates in real time, hence introducing virtually no latency to the overall system.

Furthermore, as detailed in the Methods section, the AIMC classifier, though not optimized for this task, consumes ~78 µJ per inference with a sub-ms delay (348 µs) compared to ~10 ms of digital state-of-the-art LSTM-based classifiers using filter banks for feature extraction[65]. For comparison, our system's overall energy-delay product – including the interface – outperforms the state-of-the-art digital KWS chip in 22 nm technology by more than 8.5×[66] (Suppl. Info. Table II). DNPUs and AIMC have the advantage of high integration density, *i.e.,* weight or computation capacity per area[67], compared to other CMOS technologies. Moreover, the low power consumption of DNPUs allows for bonding an AIMC to the DNPUs chip through 3D heterogenous integration, enabling a compact computing system.

As every DNPU is unique, DNPU-based feature extraction will always require the subsequent classifier to be trained accordingly. For real-life applications, it will be practical if this training is performed online in an adaptive fashion. While the inability to copy the parameters from one system to another is a disadvantage, it could be an asset for applications where safety and privacy are critical. Presently, the DNPU and AIMC layers are still physically separated. However, backend-of-the-line and heterogeneous integration[68] should allow for integration of DNPUs and AIMC on the same chip. Also, a single physical DNPU has been time-multiplexed in this work. To fully exploit our platform's potential, we need to scale up DNPU circuits in mainstream CMOS technology and use parallel multi-channel feature extraction. Although we have benchmarked our method for a speech-recognition task, we expect it to be generally applicable to temporal signal processing at the edge, such as video or EEG/ECG data classification. Our results demonstrate that combining two *in-materia* computing hardware platforms for feature extraction and classification can simultaneously address efficiency, accuracy, and compactness, paving the way towards *postconventional* heterogeneous computing systems for sustainable edge computing.

**Online content**

Any methods, additional references, Nature Research reporting summaries, source data, extended data, supplementary information, acknowledgements, peer review information; details of author contributions and competing interests; and statements of data and code availability are available at https://doi.org/xxx.

# Methods

## DNPU fabrication

A lightly n-doped silicon wafer (resistivity $\rho$~5 $\Omega\cdot$cm) is cleaned and heated for 4 hours in a furnace at 1,100 °C for dry oxidation, producing a 280 nm thick $SiO_2$ layer. Photolithography and chemical etching are used to selectively remove the silicon oxide. A second, 35 nm $SiO_2$ layer is needed for the desired dopant concentration. Ion implantation of $B^+$ ions is performed at 9 keV with a dose of $3.5\cdot10^{14}$ $cm^{-2}$. After implantation, rapid thermal annealing (1,050 °C for 7 s) is carried out to activate the dopants. The second oxide layer is removed by buffered hydrofluoric acid (1:7; 45 seconds) and then the wafer is diced into 1 × 1 cm pieces. E-beam lithography and e-beam evaporation are used, respectively, for creating the (1.5 nm Ti/25 nm Pd) electrodes. Finally, reactive ion etching ($CHF_3$ and $O_2$, (25:5)) is used to etch (30-40 nm) the silicon until the desired dopant concentration is obtained.

## DNPU measurement circuitry

We use a National Instruments (NI) C-series voltage output module (NI-9264) to apply input and control voltages to the DNPU. The NI-9264 is a 16-bit digital-to-analogue converter (DAC) with a slew rate of 4 V/μs and a settling time of 15 μs for a 100-pF capacitive load and 1 V step. As shown in Fig. 2a, a small parasitic capacitance (~10-100 pF) to ground is present at the DNPU output. In contrast to our previous work[3,4], we do not measure the DNPU output current but the output voltage, without amplification. In Refs. [3,4,48], the device output was virtually grounded by the operational amplifier used for current-to-voltage conversion (Extended Data Fig. 1). Thus, the external capacitance was essentially short-circuited to ground, and no time dynamics was observed. In the present study, we directly measure and digitize the DNPU output voltage with the NI C-series voltage input module (NI-9223; input impedance >1 G$\Omega$). A large input impedance, *i.e.*, more than ten times the DNPU resistance, is necessary to ensure that the time dynamics of the DNPU circuit is measurable.

## DNPU optimization

The DNPU control electrodes are used to tune the functionality for both the linear and nonlinear operation regimes. Applying control voltages ≳500 mV pushes the DNPU into its linear regime. Furthermore, higher control voltages make the device more conductive, leading to a faster discharge of the external capacitor and, thus, a smaller time constant. In this work, we randomly choose control voltages between -0.4 V and 0.4 V except for the end-to-end training of neural networks with DNPUs in the loop (see Suppl. Info. Note 1). For electrodes directly next to the output, we reduce this range by a factor of two because these control voltages have a stronger influence on the output voltage.

**DNPU static power measurement**

To estimate the DNPU energy efficiency, we measured the static power consumption, $P_\text{static}$, for ten different sets of random control voltages and averaged the results. In every configuration, a constant DC voltage is applied to each electrode, and the resulting current through every electrode is measured sequentially using a Keithley 236 source measure unit (SMU). $P_\text{static}$ is calculated according to

$$P_\text{static} = \sum_{k=0}^{N-1} V_i I_i,$$

where $N = 8$ is the number of electrodes of the device.

As illustrated in Extended Data Fig. 2, the average static power consumption $<P_\text{static}>$ of the measured DNPU is ~1.9 nW. For an estimate of the DNPU power efficiency, we use a conservative value of 5 nW, leading to 320 nW for 64 DNPUs in parallel, which is ~3× lower than realized with analogue filter banks reported in Ref. [35]. However, it is noteworthy to emphasise that the advantage of DNPU preprocessing extends beyond this improvement by simplifying the classification step, as illustrated in Fig. 2.

**TI-46-Word spoken digit dataset**

The audio fragments of spoken digits are obtained from the TI-46-Word dataset, available at https://catalog.ldc.upenn.edu/LDC93S9. To reduce the measurement time, we use the female subset, which contains a total of 2,075 clean utterances from 8 female speakers, covering the digits 0 to 9. The audio samples have been amplified to an amplitude range of -0.75 V to 0.75 V to match the DNPU input range and trimmed to minimize the silent parts by removing data points smaller than 50 mV (again for reducing measurement time). We used stratified randomized split to divide the dataset into train (90%) and test (10%) subsets.

**Google Speech Commands dataset**

The Google Speech Commands (GSC) dataset (available at https://www.tensorflow.org/datasets/catalog/speech_commands), is an open-source dataset containing 65,000 one-second audio recordings spoken by over 1,800 speakers. GSC is commonly used to evaluate keyword spotting systems (KWS) designed to minimise false-positive detections. While the dataset comprises thousands of audio recordings, to reduce our measurement time, we selected a subset of 6,000 recordings (100 minutes of audio, total $64 \times 100 \approx 106$ hours of measurement), comprising 200 recordings per class (total 30 classes). No preprocessing, such as trimming silence or normalising data, was applied to this subset. The dataset was divided into training (90%) and testing (10%) sets to assess the performance of our system.

**Software-based feedforward-neural-network training and inference**

To evaluate the DNPU performance in reducing the classification complexity, we combined the DNPU preprocessing with two shallow ANNs: (1) a 1-layer feedforward, and (2) a 1-layer convolutional neural network. We trained these two models for the TI-46-Word spoken digits dataset with both the original (raw) dataset and the 32-channel DNPU-preprocessed data. For all evaluations, we used the AdamW optimizer with a learning rate of $10^{-3}$ and a weight decay of $10^{-5}$ and trained the network for 200 epochs.

- **Linear layer with the original dataset.** Each digit (0 to 9) in the dataset consists of an audio signal of 1 second length sampled with a 12.5 kS/s rate. Thus, 12,500 samples have to be mapped into one of ten classes. The linear layer, therefore, has $12{,}500 \times 10 = 125{,}000$ learnable parameters followed by 10 log-sigmoid functions.
- **Linear layer with the DNPU preprocessed data.** A 10-channel DNPU preprocessing layer with a down-sampling rate of 10× converts an audio signal with a shape of 12,500×1 into 1,250×10. Then, the linear layer with 1,250×10×10 = 125,000 learnable parameters is trained. This model gives ~57% accuracy, which is 2% point less than the 32-channel result reported in Fig. 3.

- **CNN with the original dataset.** The CNN model contains a 1-D convolution layer with one input channel and 32 output channels, kernel size of 8, with a stride of 1, followed by a linear layer and log-sigmoid activation functions mapping the output of the convolution layer into ten classes. The 1-layer CNN with 32 input and output channels has ~4.5 k learnable paramters.

- **CNNs with the DNPU-processed data.** The CNN models used with DNPU preprocessed data contain one (or two) convolution layers with 16/32/64 input channels and 32 output channels followed by a linear layer. Similar to the previous model, we used a kernel size of 8 with a stride of 1 for each convolution kernel. The 1-layer CNN with 16, 32, and 64 channels has ~4.5k, ~8.6k, and ~16.9k learnable parameters, respectively.

**AIMC CNN model development**

We implemented two CNN models for classification of the TI-64-Word spoken digits dataset on the AIMC chip with 2- and 3-layer convolutional layers, trained with 32 and 64 channels of DNPU measurement data, respectively. Extended Data Fig. 3 illustrates the architecture of the 3-layer convolution layer with 64 DNPU channels (~65k learnable parameters). The first AIMC convolution layer receives the data from the DNPU with a dimension of $64 \times 1,250$. To implement this layer with a kernel size of 8, $64 \times 8 = 512$ crossbar rows are required. To optimize crossbar array resource utilization, this layer has 96 output channels. Thus, in total, 512 rows and 96 columns of the AIMC chip are utilized (Fig. 4c) to implement this layer. The second and third convolution layers both have a kernel size of 3. Considering the 96 output channels, each layer requires $96 \times 3 = 288$ crossbar rows (Fig. 4c). Finally, the fully connected layer is a $36 \times 10$ feedforward layer.

**AIMC training & inference**

The AIMC training, done in software, consists of two phases: a full-precision phase and a re-training phase, each performed for 200 epochs. The re-training phase is performed to make the classifier robust to weight noise arising from the non-ideality of the PCM devices and the 8-bit input-quantization. During this second phase, we implement two steps: (1) in every forward pass, random Gaussian noise with a magnitude equalling 12% of the maximum weight is added to each layer of the network, as well as Gaussian noise with standard deviation 0.1 is added to the output of every matrix vector multiplication (MVM) in order to make the model more robust to noise, and (2) after each training batch, weights and biases are clipped to $1.5 \times \sigma_W$ implementing the low-bit quantization, where $\sigma_W$ is the standard deviation of the distribution of weights.

**System-level efficiency analysis**

The 6-layer CNN model for the GSC dataset, implemented on the IBM HERMES project chip, possesses ~470k learnable parameters and requires 120M MAC operations per DNPU-preprocessed audio recording (all audio recordings have a duration of 1 second). Upon deployment on the AIMC chip, the model occupies 18 out of the available 64 cores (28% of the total number of cores), as depicted in Fig. 4e. Since the present chip is not designed and optimized for the studied tasks, but rather serves a general purpose, in each core, some memristive devices remain unutilized causing the efficiency to drop.

In this regard, it is necessary to mention that we use measurement reports from Ref. [5] when the chip operates in one-phase read mode, although the reported inference accuracies are for the four-phase mode. The latter approach reduces the chip's maximum throughput and energy efficiency by ~4×, while accounting for circuit and device nonidealities. Our decision to report the results based on the one-phase read mode is recently supported by the literature[70], as evidenced by the experimental demonstration of a new analogue/digital calibration procedure on the same IBM HERMES project chip. This procedure has been shown to achieve comparable high-precision computations in the one-phase read mode as those achieve in the four-phase model.

Convolution layers 0 to 3 in Fig. 4d of the main text require 1977, 492, 121, and 28 MVMs per number of occupied cores, respectively. Therefore, the total number of MVMs (including two fully connected layers) is $\sum_{l=0}^{5} MVMs_l \times$ num_cores $= 5,861$. The IBM HERMES project chip consumes 0.86 µJ at full utilization (for all 64 cores) for MVM operations with a delay of 133 ns. Consequently, the classifier model consumes $\frac{5,861}{64} \times 0.86$ µJ $= 78.7$ µJ. ). Similarly, the end-to-end latency can be calculated as $\sum_{l=0}^{5} MVMs_l \times 133$ ns $= 2,619 \times 133$ ns $= 348.3$ µs. Note that a layer (weight) matrix is typically partitioned into submatrices to be fitted on the AIMC crossbar core[71]. In our calculations, we assume that these submatrices are mapped to different cores, and therefore, the partial block-wise MVMs are executed in parallel.

Our evaluation approach stands on conservative side for MVM energy consumption; for instance, we assume energy consumption of one core for linear layers with 17,152 learnable parameters (out of 262,144 memristive devices of a core, which is only 6.5% utilization). However, we assume negligible energy consumption due to (local) digital processing, which rounds for ~7% of the total energy consumption (28% core utilization × 27% LDPU's part out of total static power consumption). Further, due to batch-norm and max-pooling layers, we buffered MVM results of each layer on memory, which introduces additional delay to the computations. However, for rea-world tasks, we can avoid CNNs but rather use large MLP layer or RNNs instead.

In Suppl. Info. Note 9, we have compared these results (summarized in Suppl. Table II) with the state-of-the-art approaches including digital and AIMC based solutions.

71      Lammie, C. et al. Lionheart: A layer-based mapping framework for heterogeneous systems with analog in-memory computing tiles. *IEEE Trans. Emerg. Topics Comput.* (2025).

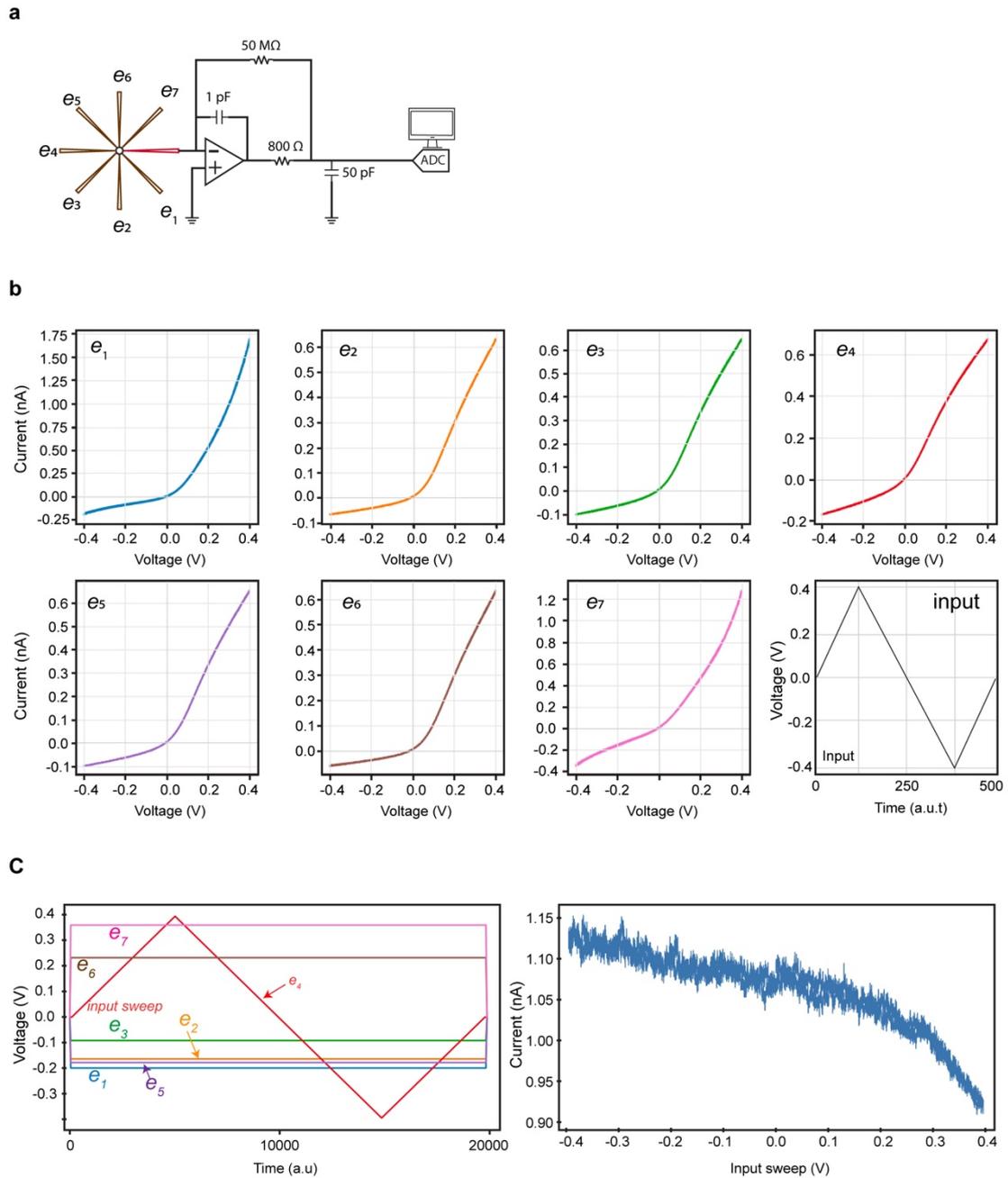

**Extended Data Fig. 1. Voltage-in current-out (static) characterization of a boron-doped DNPU. a,** Transimpedance (voltage-to-current) amplifier circuit to characterize the device. **b,** Output current of each electrode ($e_i$) in response to a voltage sweep input from -400 mV to 400 mV, while other electrodes are at zero volt. **c,** (left panel) The input electrode ($e_4$) receives an input sweep with 400 mV magnitude (shown in red) while the rest of the electrodes are at a constant voltage after an upward ramp from 0 V. (right panel) Output current of the DNPU demonstrating negative differential resistance (NDR).

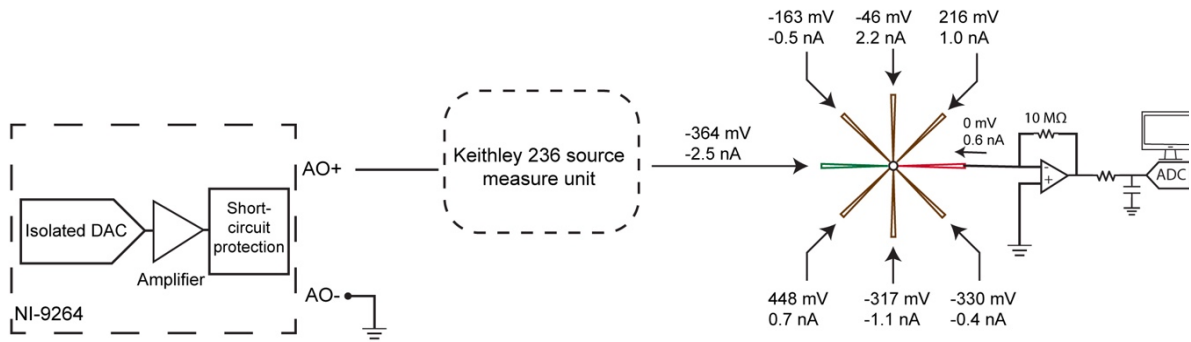

$$\langle P_{static} \rangle = \Sigma V_i I_i$$

$$= 1.9 \pm 0.5 \text{ nW}$$

**Extended Data Fig. 2. Schematic of power measurement circuit and an example of measured data.** A digital-to-analogue converter (DAC), source measure unit (SMU), and the DNPU are connected in series. For each electrode, the specified voltage is applied by the DAC, and the current is measured with a source measure unit. The measurement is repeated sequentially for each electrode.

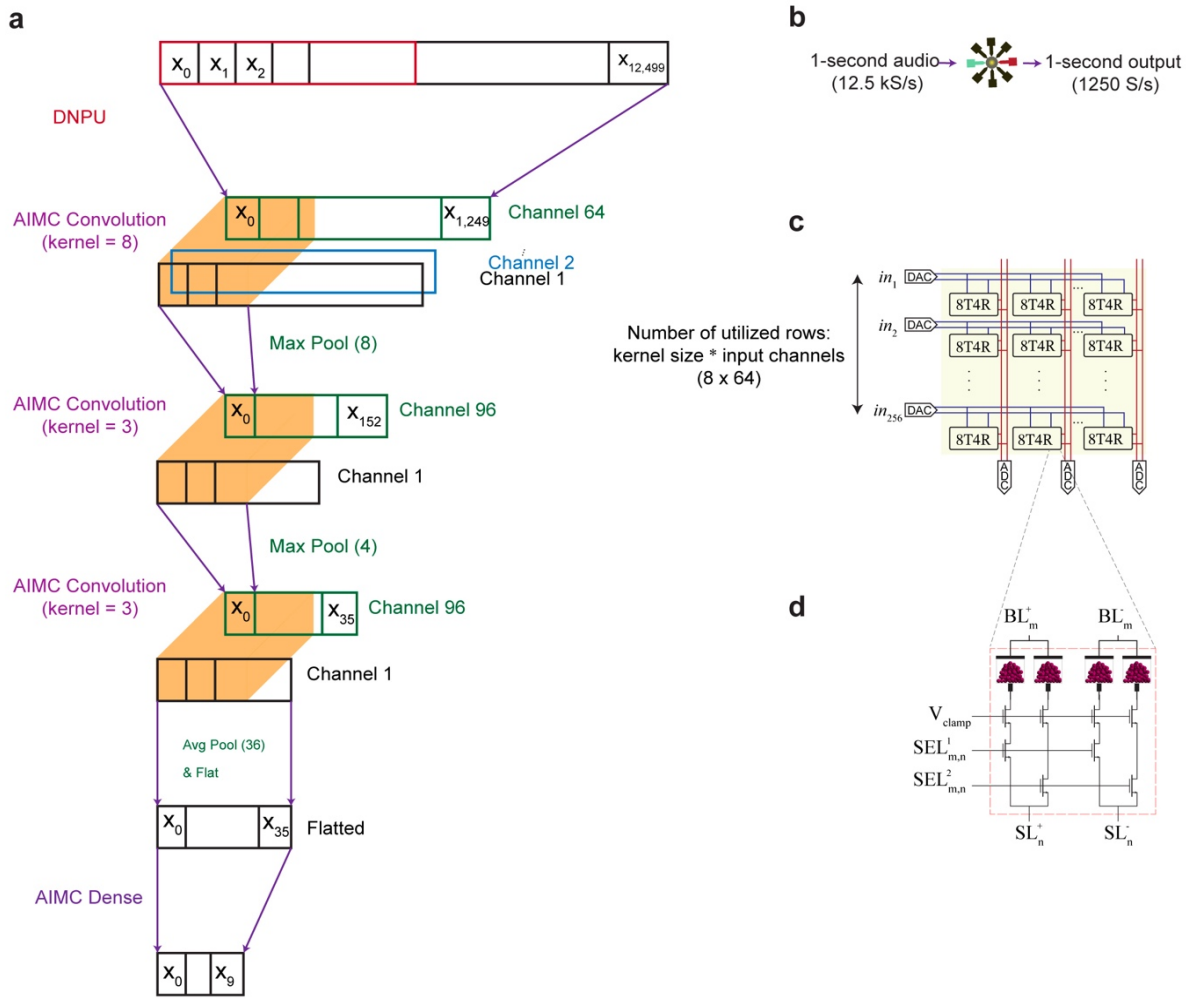

**Extended Data Fig. 3. Hybrid neural network architecture realized with dopant network processing unit (DNPU) and the IBM HERMES project analogue in-memory computing chip (AIMC). a**, Network structure. The input to the network is a 1-second audio signal sampled at 12.5 kS/s. The DNPU preprocessing produces 64 vectors of data each with 1,250 samples. Three convolution layers with kernel sizes of 8, 3 and 3 followed by a fully connected layer map the DNPU-processed signal into 10 classes. **b**, The DNPU output is downsampled by a factor of 10 by setting the ADC sampling rate to average every 10 points (oversampling). **c**, The AIMC chip consists of 256 by 256 unit cells. To implement a convolution layer, the crossbar must have a size that includes the kernel size (8 in this example) multiplied by the number of channels (64) in rows. **d**, Each synaptic unit cell of the AIMC chip comprises 4 phase-change-memory devices and 8 transistors (8T4R) organized in a differential configuration to allow for negative weights.

## Data availability

Data are available from the corresponding author upon reasonable request.

## Code availability

The simulation code and data measured from the DNPUs are publicly available on https://github.com/Mamrez/speech-recognition.

**Acknowledgments**

We thank M. H. Siekman and J. G. M. Sanderink for technical support, T. Chen, P. A. Bobbert, U. Alegre-Ibarra, and R. J. C. Cool for stimulating discussions. We acknowledge financial support from Toyota Motor Europe, the Dutch Research Council (NWO) HTSM Grant No. 16237, from the HYBRAIN project funded by the European Union's Horizon Europe research and innovation programme under Grant Agreement No 101046878. This work was further funded by the Deutsche Forschungsgemeinschaft (DFG, German Research Foundation) – SFB 1459/2 2025 – 433682494. This work was also supported by the IBM research AI Hardware Center.

**Author contributions**

M.Z. and W.G.v.d.W. designed the experiments. L.C. fabricated the DNPU devices. M.Z. and J.B. performed the measurements and simulations. J.B. ported the trained CNN models on the IBM HERMES project chip. All authors discussed the data. M.Z. and W.G.v.d.W. wrote the manuscript and all the authors contributed to revisions. W.G.v.d.W., A.S., and G.S.S. conceived the project. W.G.v.d.W. supervised the project.

**Competing interests**

The authors declare no competing interests.

**Supplementary information** is available for this paper at https://doi.org/xxx

**Correspondence and requests for materials** should be addressed to W.G.v.d.W.

**Peer review information**

**Reprints and permissions information** is available at http://www.nature.com/reprints.